\documentclass[prb,aps,amsmath,amssymb,twocolumn,superscriptaddress]{revtex4-2}
\usepackage[colorlinks=true,urlcolor=blue,bookmarks=true,citecolor=magenta,breaklinks=true,pdftex]{hyperref}

\usepackage{graphicx}
\usepackage{dcolumn}
\usepackage{bm}
\usepackage{etoolbox}
\usepackage{wrapfig}
\usepackage{lipsum}
\usepackage{latexsym}
\usepackage{xtab}
\usepackage{afterpage}
\usepackage{threeparttablex}
\usepackage{xcolor}
\usepackage{multirow}
\usepackage{hhline}
\usepackage{booktabs}
\usepackage{makecell}
\usepackage{cleveref}
\usepackage{orcidlink}
\usepackage{comment}

\usepackage{subfig}
\usepackage[LGR,T1]{fontenc}
\usepackage[utf8]{inputenc}
\DeclareSymbolFont{upgreek}{LGR}{cmr}{m}{n}
\SetSymbolFont{upgreek}{bold}{LGR}{cmr}{bx}{n}
\DeclareMathSymbol{\Iota}{\mathord}{upgreek}{`I}
\newcommand{\overlap}[2]{\left\langle {#1} \middle\vert {#2} \right\rangle}
\newcommand{\matelem}[3]{\left\langle {#1} \middle\vert {#2} \middle\vert {#3} \right\rangle}

\definecolor{darkblue}{rgb}{0.0,0.0,0.75}
\newcommand{\update}[1]{{ \color{black} #1} }

\renewcommand\Im{\operatorname{Im}}

\begin{document}

\title{Real-time exciton dynamics in two-dimensional materials under ultrashort laser pulses}

\author{Dmitry Tumakov\orcidlink{0000-0003-0708-2427}}
\email{dm.tumakov@gmail.com}
\affiliation{Department of Physics and CSMB, Humboldt-Universit\"{a}t zu Berlin, Zum Gro{\ss}en Windkanal 2, Berlin 12489, Germany}
\affiliation{I. Institute for Theoretical Physics, University of Hamburg, Notkestra{\ss}e 9-11, Hamburg 22607, Germany}

\author{Daria Popova-Gorelova\orcidlink{0000-0002-3036-0467}}
\email{daria.gorelova@b-tu.de}
\affiliation{Institute of Physics, Brandenburg University of Technology Cottbus-Senftenberg, Erich-Weinert-Stra{\ss}e 1, Cottbus 03046, Germany}
\affiliation{I. Institute for Theoretical Physics, University of Hamburg, Notkestra{\ss}e 9-11, Hamburg 22607, Germany}

\begin{abstract}
  The optical response of two-dimensional materials is often significantly impacted by excitonic effects due to the reduced screening of attractive Coulomb interactions in low-dimensional systems. Accurate modeling of exciton formation and real-time dynamics is essential to understanding their ultrafast optical properties. In this study, we theoretically investigate the exciton dynamics in a two-dimensional hexagonal boron nitride (h-BN) and a germanium sulfide (GeS) monolayers exposed to an ultrashort laser pulse. We analyze the system's response to the external field in one- and two-photon excitation regimes. For our calculations, we combine a state-of-the-art ab initio approach to study exciton dynamics with a highly precise numerical scheme. We incorporate electron-hole interactions through a non-local self-energy operator derived from the many-body perturbation theory (MBPT) within the time-dependent adiabatic $GW$ (TD-a$GW$) approximation. We implement this approach using the full-electron LAPW+lo method in the all-electron \texttt{exciting} package.  Our results elucidate the role of many-body effects in shaping ultrafast excitonic processes in two-dimensional materials, contributing to the fundamental understanding necessary for optoelectronic and photonic applications.
\end{abstract}

\maketitle

%%%%%%%%%%%%%%%%%%%%%%%%
\section{Introduction}\label{section:intro}
%%%%%%%%%%%%%%%%%%%%%%%%
Due to the rapid development of the laser facilities, particularly their ability to generate ultrashort intense pulses~\cite{huang_innovations_2021}, the study of real-time out-of-equilibrium electron dynamics has received significant attention~\cite{zong_natrev_2023, chergui_natrev_2023, cavaletto_natrev_2024}. Today, it is possible to observe electron dynamics on their natural attosecond timescale. In many light-matter interaction experiments with solids, excitonic effects, which stem from the strong Coulomb interaction between an electron in the conduction band (CB) and a hole in the valence band (VB), can play an essential role in charge dynamics. These effects are particularly prominent in non-metallic two-dimensional (2D) materials, where the attractive Coulomb interaction is less suppressed by screening. This can lead to excitonic binding energies of about 1 eV~\cite{wang_revmodphys_2018}, which makes excitons stable at room temperature. Despite extensive research, the fundamental mechanisms underlying exciton formation and dynamics, particularly in the nonlinear response regime, are still not well understood. \\

Understanding these complex excitonic dynamics requires advanced experimental and theoretical approaches. Among these approaches, ultrafast spectroscopic techniques have emerged as powerful tools for probing the real-time behavior of electrons in materials.  Most of the commonly used ultrafast techniques rely on a pump-probe scheme. A pump laser pulse, with a photon energy ranging from a tenth of an eV to several eV excites a target, initiating transient charge dynamics that are subsequently probed by a second pulse. Techniques such as attosecond transient absorption spectroscopy (ATAS), X-ray absorption spectroscopy (XAS), angle-resolved photoemission spectroscopy (ARPES), and transient absorption spectroscopy (TR-abs) can be implemented in a time-resolved manner by controlling the pump-probe delay~\cite{stolow_chemr_2004}. Another widely used technique is high-frequency light produced by pump-driven charge dynamics, known as solid-state high harmonic generation (sHHG)~\cite{ghimire_nat_2019}. It has been shown that excitonic effects can significantly impact sHHG, if present~\cite{molinero_sciadv_2024}. A pump-probe setup inherently involves a nonlinear response due to interaction with two light pulses. Even a pump pulse itself can be intense enough to induce nonlinear effects. Therefore, a comprehensive theoretical description of these experiments should ideally account for light-matter interaction beyond the first order.\\

  Excitonic effects can be incorporated using various real-time methods such as the propagation of the nonequilibrium Green's function~\cite{attaccalite_prb_2011, stefanucci_2013, perfetto_prb_2015, perfetto_jct_2019, ridolfi_prb_2020, chan_pnas_2021, sangalli_prm_2021, perfetto_prl_2022, hou_npjcm_2025, marek_jctc_2025}, non-adiabatic RT-TDDFT~\cite{sander_jcp_2017, pemmaraju_njp_2020, sun_prl_2021, williams_jctc_2021}, real-time propagation of the Bethe-Salpeter equation (BSE)~\cite{jiang_sciadv_2021}, semiconductor Bloch equations~\cite{cistaro_jctc_2023, malakhov_commphys_2024, molinero_sciadv_2024}, and time-dependent Hartree-Fock (TDHF) simulations~\cite{jensen_pra_2024}. In the frequency space, a target's nonlinear response can be studied using the lowest-nonzero-order perturbation theory with the solutions of the BSE~\cite{riefer_prb_2017, vorwerk_prr_2020, sangalli_prb_2023, farahani_prb_2024, ruan_nanolett_2024}. While these methods provide complementary routes to describe excitonic effects, they differ significantly in computational cost, scalability, and ease of integration with existing electronic-structure frameworks. \\

To account for excitonic effects alongside the nonlinear response to a laser field, we adopt the effective Schr\"{o}dinger equation approach, suggested in Ref.~\cite{attaccalite_prb_2013}. We choose this framework because it provides a computationally efficient real-time scheme that retains the key many-body physics of the BSE while avoiding the explicit propagation of two-particle Green's functions or the full BSE kernel in time. This method utilizes nonequilibrium Berry-phase coupling with an external field~\cite{souza_prb_2004} following the ``modern theory of polarization''~\cite{king_prb_1993, resta_revmodphys_1994}. It incorporates dynamical electron-hole interaction effects via the many-body perturbation theory (MBPT)-derived self-energy operator~\cite{attaccalite_prb_2011, attaccalite_prb_2013,pionteck_scipost_2025}. This technique, which combines the adaptability of a real-time approach
with the natural ability of MBPT to capture electron correlation, has been successfully used for nonlinear response calculations~\cite{attaccalite_prb_2013, gruening_prb_2014, attaccalite_prb_2017, attaccalite_prb_2018, pionteck_arxiv_2025}. Its numerical implementation is available in the open-source {YAMBO} code, which uses a plane wave basis and pseudopotential approach~\cite{marini_cpc_2009, sangalli_jphys_condmat_2019}. \\

Here, we implement this functionality in the real-time time-dependent density functional theory (RT-TDDFT) module of the all-electron \texttt{exciting} package~\cite{gulans_jphys_condmat_2014, pela_electrstruct_2021, exciting_2026}, which uses the linearized augmented plane-wave plus local orbitals (LAPW+lo) basis set. This basis set enables an accurate description of both delocalized and localized electronic states. LAPW+lo provides a highly precise numerical scheme for solving the Kohn-Sham equations \cite{gulans_jphys_condmat_2014}. In addition to improved accuracy, it allows for the future description of ultrafast XAS experiments due to its accurate treatment of core electrons. \\

We present a detailed theoretical study of the real-time charge dynamics of a 2D hexagonal boron nitride (h-BN) monolayer exposed to an ultrashort laser pulse using our highly accurate implementation. This prototype 2D material is often nicknamed as ``white graphene'' due to its similar lattice structure~\cite{liu_nat_2013}. This insulator with a wide gap of $\sim$6-7 eV~\cite{ba_scirep_2017, elias_natcom_2019, roman_2dmats_2021} has been extensively studied both theoretically and experimentally over the decades due to its high thermal and chemical stability, as well as its optoelectrical properties~\cite{cassabois_natphot_2016, galvani_prb_2016, roldan_csr_2017, wickramaratne_jpc_2018, ferreira_josab_2019, roy_advmat_2021, rousseau_nanolett_2021, kirchhoff_prb_2022, zhang_prl_2022, molinero_sciadv_2024}. We study a resonant exciton formation by one- and two-photon absorption and describe the resulting dynamics. Another system we consider is a germanium monosulfide (GeS) monolayer, which has a moderate band gap of $\sim$2.3 eV and a large predicted carrier mobility~\cite{li_jmcc_2016}, making it promising for various photoelectric applications~\cite{li_jmcc_2016, dien_prb_2023, esteve_npj_2025}. \\

The paper is organized as follows. Section~\ref{section:theory} 
describes the theoretical method used in detail. Section~\ref{section:implementation} provides details on the numerical implementation. The results of our calculations are presented
in Section~\ref{section:results}. Section~\ref{section:summary} contains the concluding remarks.\\

%%%%%%%%%%%%%%%%%%%%%%%%
\section{Theory}\label{section:theory}
%%%%%%%%%%%%%%%%%%%%%%%%
In this section, we summarize the real-time approach developed in Refs.~\cite{souza_prb_2004, attaccalite_prb_2013}, which we implement in the \texttt{exciting} code, and employ in our calculations. \\

\subsection{Real-time equations of motion}
The dynamical macroscopic polarization along the lattice vector $\bm{a}_{\alpha}$ ($\alpha$ = 1, 2, 3) obtained within the modern theory of polarization~\cite{king_prb_1993, resta_revmodphys_1994, souza_prb_2004} is given by
\begin{equation}\label{eq:polarization}
    \bm{P}_{\alpha} = \frac{f \bm{a}_{\alpha}}{2 \pi V N_{\bm{k}_{\alpha}^{\perp}}} \sum_{\bm{k}_{\alpha}^{\perp}} {\rm Im \, ln} \, \prod_{i = 1}^{N_{\bm{k}_{\alpha}} - 1} {\rm det} \, S^{\bm{k}_i, \bm{k}_{i \alpha}^+},
\end{equation}
\begin{equation}\label{eq:tds}
    S^{\bm{k}, \bm{k}_{\alpha}^{\pm}}_{mn} = \overlap{v_{\bm{k}m}}{v_{\bm{k}_{\alpha}^{\pm} n}}.
\end{equation}
Here, $V$ is the unit cell volume, $f$ is the spin degeneracy factor equal to two in this study, and $N_{\bm{k}_{\alpha}}$ is the number of equally distributed $\bm{k}$ points along the reciprocal lattice vector $\bm{b}_{\alpha}$, such that $\bm{a}_{\alpha} \cdot \bm{b}_{\alpha} = 2 \pi$. $N_{\bm{k}_{\alpha}^{\perp}}$ is the number of $\bm{k}$ points in the plane orthogonal (in lattice coordinates) to $\bm{b}_{\alpha}$, $\bm{k}_{\alpha}^{\pm} = \bm{k} \pm \Delta \bm{k}_{\alpha}$, and $\Delta \bm{k}_{\alpha} = \bm{b}_{\alpha} / N_{\bm{k}_{\alpha}} $. We use atomic units (a.u.), $\hbar = m_{e} = -e = 1$, for this and the following equations. The time-dependent functions $v_{\bm{k}m}$ are the spatially periodic parts of the time-dependent Kohn-Sham-Bloch states $\phi_{\bm{k}m} (\bm{r}, t)$:
\begin{equation}
    \phi_{\bm{k}m} (\bm{r}, t) = e^{i \bm{k} \cdot \bm{r}} v_{\bm{k}m} (\bm{r}, t), \; v_{\bm{k}m} (\bm{r} + \bm{R}, t) = v_{\bm{k}m} (\bm{r}, t),
\end{equation}
where $\bm{R}$ is a Bravais lattice vector. Matrix $S^{\bm{k}_i, \bm{k}_{i \alpha}^+}$ is defined in Eq.~\eqref{eq:tds} for a fixed spin projection. We restrict the consideration to the spin-symmetric case. \\

The Lagrangian for an infinite system with a uniform grid of $N_{\bm{k}}$ $\bm{k}$-points in the Brilloin zone (BZ) in the presence of a classical homogeneous electric field $\bm{E}(t)$ reads
\begin{equation}\label{eq:lagrangian}
    L = \frac{1}{N_{\bm{k}}} \sum_{n, \bm{k}} \left[ i \overlap{v_{\bm{k}n}}{\dot{v_{\bm{k}n}}} - \matelem{v_{\bm{k}n}}{H^0_{\bm{k}}}{v_{\bm{k}n}} \right] + V \bm{E}(t) \cdot \bm{P}(t),
\end{equation}
where $\bm{P}(t)$ is the total macroscopic polarization, $\bm{P}(t) = \sum_\alpha \bm{P}_\alpha(t)$. 
The Euler-Lagrange equations for the Lagrangian~\eqref{eq:lagrangian} can be recast as the following equations of motion for the periodic parts of the Bloch states:
\begin{equation}\label{eq:dynamics}
    i \frac{d}{dt}  v_{\bm{k} m}(t)  = \left[ H^0_{\bm{k}} + V^{\rm ext}_{\bm{k}} (t) \right] v_{\bm{k} m}(t),
\end{equation}
\begin{equation}
    H^0_{\bm{k}} = e^{-i \bm{k} \cdot \bm{r}'} H^0 e^{i \bm{k} \cdot \bm{r}},
\end{equation}
where $H^0$ is the field-free system's Hamiltonian, $\bm{r}'$ and $\bm{r}$ correspond to the coordinates of the bra- and ket-states respectively, and $V^{\rm ext}_{\bm{k}} (t)$ is the operator describing the interaction with the external electromagnetic field:
\begin{equation}\label{eq:vext}
    V^{\rm ext}_{\bm{k}} (t) =  w_{\bm{k}}(\bm{E}(t)) + w_{\bm{k}}^{\dagger}(\bm{E}(t)),
\end{equation}
\begin{equation}\label{eq:wk}
    w_{\bm{k}}(\bm{E}(t)) = \frac{i}{4 \pi} \sum_m \sum_{\alpha = 1}^3 (\bm{a}_{\alpha} \cdot \bm{E}(t) ) N_{\bm{k}_{\alpha}} \sum_{\sigma = \pm} \sigma \left| \tilde{v}_{\bm{k}_{\alpha}^{\sigma} m} \right\rangle \left \langle v_{\bm{k} m} \right |,
\end{equation}
where 
\begin{equation}
    \tilde{v}_{\bm{k}_{\alpha}^{\pm} m}  = \sum_n \left[ S^{\bm{k}, \bm{k}_{\alpha}^{\pm}} \right]^{-1}_{nm} v_{\bm{k}_{\alpha}^{\pm} n}.
\end{equation}
We note that for any $m$, $w_{\bm{k}}^{\dagger} (\bm{E}(t)) \left| v_{\bm{k} m} (t) \right \rangle = 0$, and the second term in Eq.~\eqref{eq:vext} is included only to ensure that the resulting operator is Hermitian. It can be shown that the one-particle density matrix constructed with the solutions of Eq.~\eqref{eq:dynamics} as
\begin{equation}\label{eq:density}
    \rho (\bm{r}, \bm{r}'; t) = \Omega^{-1} \sum_{m = 1}^{n_{\rm occupied}} \int_{\rm BZ} d \bm{k} \phi_{\bm{k}m} (\bm{r}, t) \phi^*_{\bm{k}m} (\bm{r}', t)
\end{equation}
evolves according to the Liouville–von Neumann equation~\cite{souza_prb_2004}. Here, $\Omega$ is the BZ volume and the index $m$ runs over $n_{\rm occupied}$ initially occupied states in the valence band (VB). \\

With the time-dependent functions $v_{\bm{k} m}(t)$ and, thus, the time-dependent overlap matrix~\eqref{eq:tds}, the macroscopic polarization can be obtained using Eq.~\eqref{eq:polarization}. After time propagation, macroscopic response properties can be extracted from the time-dependent polarization vector $\bm{P}(t)$. \\

We assume the dipole approximation for the classical electromagnetic field $\bm{E}(t)$, which is switched on at some time $t_0 \geqslant 0$. Within this approximation, the electric field is spatially uniform, thus only vertical optical transitions are possible, and no magnetic field effects are present. This approximation is well established for the moderate field strengths and relatively long wavelengths used in our calculations. \\

All post excitation relaxation  processes such as carrier cooling, scattering, transport, and recombination~\cite{wheeler_advmat_2013} are neglected in our study, as we focus on the first few tens of femtoseconds following the excitation. \\

\subsection{Electron-electron dynamic interaction}
We define the explicitly field-free Hamiltonian as
\begin{equation}
    H^0 \equiv H^{\rm GS} + \delta V^{\rm ee}[\rho(t)],
\end{equation}
where $\delta V^{\rm ee}$ contains a time-dependent correction to the inter-electronic interaction. The ground-state (GS) Hamiltonian reads
\begin{equation}\label{eq:gs_ham}
    H^{\rm GS} = T + V^{\rm nucl} + V^{\rm H} [\rho(0)] + V^{\rm xc}[\rho(0)] + \Delta H^{\rm QP},
\end{equation}
where $T$ is the total one-particle kinetic energy operator, $V^{\rm nucl}$ describes the interaction with static nuclei, $V^{\rm H}$ is the Hartree potential responsible for the local field effects~\cite{adler_pr_1962}, $V^{\rm xc}$ is the local KS exchange-correlation potential, and $\Delta H^{\rm QP}$ is the state-dependent scissor operator~\cite{baraff_prb_1984, onida_revmod_2002} rigidly shifting the CB on the value $\Delta E^{\rm QP}$ upwards:
\begin{equation}\label{eq:scissor}
    \Delta H^{\rm QP} = \Delta E^{\rm QP} \sum_{m \in {\rm CB}} \psi_{\bm{k} m} (\bm{r}') \psi^*_{\bm{k} m} (\bm{r}).
\end{equation}
The value of $\Delta E^{\rm QP}$ can be chosen from experimental data or results of $GW$ calculations present in literature. \update{We note that the Hartree $V^{\rm H}$ and exchange-correlation $V^{\rm xc}$ contributions only require the diagonal part of the density matrix $\rho$.} In the simulations, we compare different levels of accuracy to account for the dynamic electron-electron interaction. The first level is the independent particle approximation (IPA), with the effective KS potential being fixed to its unperturbed ground-state value:
\begin{equation}\label{eq:hamIPA}
\delta V^{\rm ee, IPA}[\rho(t)] = 0.    
\end{equation}
The electron-hole interaction effects are incorporated using the time-dependent adiabatic $GW$ (TD-a$GW$) approximation (also referred to as the time-dependent Hartree + Screened EXchange (HSEX) approximation): 
\begin{equation}\label{eq:hamFull}
\delta V^{{\rm ee, a}GW}[\rho] = V^{\rm H} [\rho(t)] - V^{\rm H} [\rho(0)] + \delta V^{\rm ee, SEX}[\rho(t)],
\end{equation}
where the self energy is obtained within the SEX (Screened EXchange)~\cite{hedin_pra_1965, hybertsen_prb_1986, strinati_smth_1988} approximation:
\update{
\begin{equation}\label{eq:hamSex}
\delta V^{\rm ee, SEX}[\rho(t)] = \Sigma^{\rm SEX}[\rho(t)] - \Sigma^{\rm SEX}[\rho(0)].
\end{equation}
}
Self energy term, which is time-dependent through the time-dependent density matrix, is defined as:
\begin{equation}
    \Sigma^{\rm SEX}[\rho(t)] = - \rho(t) W_0 (\bm{r}, \bm{r}'; \, \omega = 0), 
\end{equation}
where $W_0$ is the static (instantaneous) limit of the dynamically screened Coulomb interaction, which can be obtained in the reciprocal space:
\begin{equation}
    W_0(\bm{r}, \bm{r}') = \sum_{\bm{G} \bm{G}' \bm{q}} e^{i (\bm{G} + \bm{q}) \bm{r}} \varepsilon^{-1}_{\bm{G} \bm{G}'} (\bm{q}) v (\bm{q} + \bm{G}') (\bm{q}) e^{-i (\bm{G}' + \bm{q}) \bm{r}'},
\end{equation}
\begin{equation}
    \varepsilon_{\bm{G} \bm{G}'} (\bm{q}) = \delta_{\bm{G} \bm{G}'} - v(\bm{q} + \bm{G}) \chi_{\bm{G} \bm{G}'} (\bm{q}),
\end{equation}
where the bare Coulomb potential reads
\begin{equation}
    v (\bm{q}) = 4 \pi / q^2.
\end{equation}
The polarizability $\chi_{\bm{G} \bm{G}'} (\bm{q})$ is constructed within the random phase approximation (RPA):
\begin{equation}
    \chi_{\bm{G} \bm{G}'} (\bm{q}) = \sum_{n n' \bm{k}} M^{\bm{k} + \bm{q} \, \bm{k}}_{\bm{G} n' n } M^{\bm{k} + \bm{q} \, \bm{k}}_{\bm{G}' n' n }  \frac{f(\epsilon_{\bm{k} + \bm{q} \, n'}) - f(\epsilon_{\bm{k}n}) }{\epsilon_{\bm{k} + \bm{q} \, n'} - \epsilon_{\bm{k} n}},
\end{equation}
where $f(\epsilon_{\bm{k}n})$ are the occupation numbers, and 
\begin{equation}
    M^{\bm{k} \bm{k}'}_{\bm{G} p n }= \int d \bm{r} \psi^*_{\bm{k} p} (\bm{r}) e^{i (\bm{G} + \bm{k} - \bm{k}') \bm{r}} \psi_{\bm{k}' n} (\bm{r}).
\end{equation}
The TD-a$GW$ approximation reduces to the BSE in the linear response limit~\cite{attaccalite_prb_2011}. \\

Since we employ the length gauge for the external field coupling, non-local operators such as the scissor correction or the self-energy operator can be added directly to the Hamiltonian without any complications specific for the velocity gauge. Namely, the operators do not need to be gauge-transformed, and the velocity operator does not need to be redefined~\cite{sangalli_prb_2017}.

\subsection{Electric field strength limitations}
Despite the field coupling term being non-perturbative, there are still limitations on the peak field strength. Due to Zener tunneling, the operator~\eqref{eq:vext} becomes unbounded from below~\cite{springborg_prb_2008} at a critical field amplitude $\bm{E}^{\rm crit}$, which can be estimated as~\cite{souza_prl_2002}:
\begin{equation}
    | \bm{E}^{\rm crit} \cdot \bm{a}_{\alpha} | = \frac{E_{\rm gap}}{N_{\bm{k}_{\alpha}}},
\end{equation}
where $E_{\rm gap}$ is a fundamental gap. In all calculations presented here, the field strength does not exceed a few percent of this critical value. \update{As discussed in Ref.~\cite{souza_prb_2004}, the field which exceeds this critical value, can still be used with caution in the dynamical case.} Another limitation arises from the approximation employed for the self-energy operator: neglecting its explicit time dependence is only valid when the excitonic density $n_{\rm exc}$ remains below the Mott transition threshold~\cite{snoke_ssc_2008, perfetto_prb_2020, sangalli_prb_2023}. This threshold can be estimated using the exciton's characteristic coordinate-space size $r_{\rm exc}$ as $n_{\rm exc} \sim  r_{\rm exc}^{-d}$ with $d$ being the system's dimensionality. 

% For the lowest-lying excitons in 2D h-BN $r_{\rm exc} \lesssim 10$ unit cell sizes~\cite{galvani_prb_2016, paleari_2dmats_2018, zhang_prl_2022, malakhov_commphys_2024}, and thus $n_{\rm exc}$ is around 0.01 per cell, which is significantly higher than the values obtained in the present calculations. 

\section{Implementation details}\label{section:implementation}

 In the simulations, we expand the time-dependent states $v_{\bm{k}n}(t)$ with the set of the periodic parts of the ground state Kohn-Sham-Bloch orbitals:
 \update{
\begin{equation}\label{eq:expansion}
     v_{\bm{k} n} (t) = \sum_{m = 1}^N c_{\bm{k} mn} (t) u_{\bm{k} m},
\end{equation}
}
where the time-independent functions $u_{\bm{k} m}$ are defined by
\begin{equation}
    \psi_{\bm{k} m} (\bm{r}) = e^{i \bm{k} \cdot \bm{r}} u_{\bm{k} m} (\bm{r}), \; u_{\bm{k} m} (\bm{r} + \bm{R}) = u_{\bm{k} m} (\bm{r}),
\end{equation}
\begin{equation}\label{eq:gs_states}
    H^0 \psi_{\bm{k} m} = \epsilon_{\bm{k} m} \psi_{\bm{k} m}, \; m = 1, \dots, N.
\end{equation}
The value of $N = n_{\rm occupied} + n_{\rm empty}$ is defined by the number $n_{\rm empty}$ of initially unoccupied conduction band (CB) states included in the basis set. Substituting the expansion~\eqref{eq:expansion} into Eq.~\eqref{eq:dynamics}, we obtain the following equation for the expansion coefficients:
\begin{equation}\label{eq:coeffsEq}
    i \frac{d}{dt} c_{\bm{k} pn} (t)  = \sum_m  \left[ \matelem{u_{\bm{k}p}}{H^0_{\bm{k}} + V^{\rm ext}_{\bm{k}} (t)}{u_{\bm{k}m}} \right] c_{\bm{k} mn} (t).
\end{equation}
We set $c_{\bm{k} pn}(0) = \delta_{pn}$ for each $\bm{k}$ and $p$ as the initial condition for Eqs.~\eqref{eq:coeffsEq}. The set of matrix differential equations~\eqref{eq:coeffsEq} is evolved in time using the 
 4th-order Runge-Kutta method (RK4) available in the \texttt{exciting} package~\cite{pela_electrstruct_2021}. \\

 As the functions $v_{\bm{k}m}(t)$ are constructed and evolved numerically independently for different $\bm{k}$ points, they acquire independently a random initial phase. By construction, the operator in Eq.~\eqref{eq:vext}, which couples neighboring $\bm{k}$ points, is gauge-covariant and satisfies the Born-von Karman periodic boundary conditions~\cite{souza_prb_2004}. Eq.~\eqref{eq:wk} implies a two-point finite-difference approximation for the derivative in $\bm{k}$ space. Achieving a higher level of accuracy is possible by applying it twice, with step sizes of $\Delta \bm{k}_{\alpha}$ and $2 \Delta \bm{k}_{\alpha}$, provided there are at least five $\bm{k}$ points along the $\bm{b}_\alpha$ direction~\cite{nunes_prb_2001}:
\begin{equation}
    w_{\bm{k}} = \frac{1}{3} \left ( 4 w_{\bm{k}}|_{\Delta \bm{k}_{\alpha}} - w_{\bm{k}}|_{2 \Delta \bm{k}_{\alpha}} \right ).
\end{equation}
 
 To reduce the computational cost, we assume that $n_{\rm frozen}$ lowest-lying valence states for each $\bm{k}$ point are irrelevant for a specific process under consideration, and only propagate $n_{\rm active} = n_{\rm occupied} - n_{\rm frozen}$ states, i.e. $c_{\bm{k} mn}(t) = \delta_{mn}$ for $n = 1, \dots n_{\rm frozen}$. The results are checked for convergence both with decreasing $n_{\rm frozen}$ and increasing $n_{\rm empty}$. \\

The details of evaluation of the local operators' matrix elements in the \texttt{exciting} package can be found in Ref.~\cite{gulans_jphys_condmat_2014}. The field coupling matrix $\matelem{u_{\bm{k}p}}{V^{\rm ext}_{\bm{k}} (t)}{u_{\bm{k}m}}$ involves a time-dependent overlap between states from different $\bm{k}$ points:
\begin{equation}
    S^{\bm{k}, \bm{k}_{\alpha}^{\sigma}}_{ln} (t)  = \sum_p c^*_{\bm{k} pl} (t) G^{\bm{k} \bm{k}_{\alpha}^{\sigma}}_{pn}(t),
\end{equation}
where
\begin{equation}
    G^{\bm{k} \bm{k}_{\alpha}^{\sigma}}_{pn}(t) = \sum_{r} \overlap{u_{\bm{k} p}}{u_{\bm{k}_{\alpha}^{\sigma} r}} c_{\bm{k}_{\alpha}^{\sigma} rn}(t).
\end{equation}
This introduces real-time communication between different MPI (Message Passing Interface) processes when $\bm{k}$-grid parallelization is employed.
For all $\alpha = 1, 2, 3$ and $\sigma = \pm 1, \pm 2$, we keep the time-independent $n_{\rm occupied} \times n_{\rm occupied}$ matrix $\overlap{u_{\bm{k} p}}{u_{\bm{k}_{\alpha}^{\sigma} r}}$ available on the MPI process controlling the point $\bm{k}$. In this way, communication is reduced to an inexpensive exchange of $(n_{\rm occupied} + n_{\rm empty}) \times n_{\rm active}$ matrices of the coefficients $c_{\bm{k} rn}(t)$. The matrix 
\begin{equation}
    \overlap{u_{\bm{k} m}}{u_{\bm{k}_{\alpha}^{\sigma} n}} = \int \psi_{\bm{k} m}^* (\bm{r}) e^{-i \left(\bm{k}_{\alpha}^{\sigma} - \bm{k} \right) \bm{r}}  \psi_{\bm{k}_{\alpha}^{\sigma} + s \bm{b}_{\alpha} \, n} (\bm{r}) d \bm{r},
\end{equation}
where $s = 0, \pm 1$ depending on whether the neighboring $\bm{k}$ points lie in the same Brillouin zone, is calculated just once prior to the time propagation. 
The Hartree potential matrix $\matelem{\psi_{\bm{k}p}}{V^{\rm H} [\rho(t)]}{\psi_{\bm{k}m}}$ is evaluated in the linearized augmented planewave and local orbitals (LAPW+lo) basis set employed in the \texttt{exciting}, and then transformed to the KS basis. \\

For the efficient real-time recalculation of the dynamical interelectronic interaction potential, we adopt the approach employed in the {YAMBO} code~\cite{attaccalite_prb_2011}. First, we precalculate the four-point matrix 
\begin{equation}
    \mathcal{W}^{\bm{k} \bm{k}'}_{p m n l} = \sum_{\bm{G} \bm{G}'} W_{\bm{G} \bm{G}'} (\bm{k} - \bm{k}') M^{\bm{k} \bm{k}'}_{\bm{G} p n } \left( M^{\bm{k} \bm{k}'}_{\bm{G}' m l } \right)^*.
\end{equation}
Then, the matrix elements of the self-energy operator are obtained at each time propagation step as
\begin{equation}\label{eq:tdsex}
    \matelem{\psi_{\bm{k}p}}{ \Sigma^{\rm SEX}}{\psi_{\bm{k}m}} = - \sum_{n l \bm{k}'} \mathcal{W}^{\bm{k} \bm{k}'}_{p m n l} \sum_{s = 1}^{n_{\rm occupied}} c_{\bm{k} ns}(t) c^*_{\bm{k} ls}(t).
\end{equation}
We work within the following approximation: in the summation in Eq.~\eqref{eq:tdsex}, we only include terms where the pairs {$p$, $n$} and {$m$, $l$} lie within the same band. Omitting the remaining terms is not expected to affect our results~\cite{sangalli_prb_2023}. For the details of calculation of the matrix $\mathcal{W}^{\bm{k} \bm{k}'}_{p m n l}$, which is the direct interaction term in BSE, in \texttt{exciting}, please refer to Ref.~\cite{vorwerk_es_2019}. \update{We also refer to a recent study in Ref.~\cite{hou_npjcm_2025}, where a more efficient implementation of the real-time recalculation of the SEX self energy term in Eq.~\eqref{eq:tdsex} was reported.} \\

% Hartree term can also be recalculated fast when using the same approach:
% \begin{equation}\label{eq:tdh}
%     \matelem{\psi_{\bm{k}p}}{ V^{\rm H}}{\psi_{\bm{k}m}} =  \sum_{n l \bm{k}'} \mathcal{V}^{\bm{k} \bm{k}'}_{p m n l} \sum_{s = 1}^{n_{\rm occupied}} c_{\bm{k} ns}(t) c^*_{\bm{k} ls}(t),
% \end{equation}
% \begin{equation}
%      \mathcal{V}^{\bm{k} \bm{k}'}_{pmrl} = \sum_{\bm{G}} v^{\rm C}(\bm{G}) M^{\bm{k} \bm{k}}_{\bm{G} p m } \left( M^{\bm{k}' \bm{k}'}_{\bm{G} rl } \right)^*.
% \end{equation}

Finally, we would like to make a remark regarding the evaluation of the macroscopic polarization. Contributions from different ``strings'' (i.e., $N_{\bm{k}_{\alpha}^{\perp}}$ sets of $\bm{k}$ points along the direction $\alpha$) in Eq.~\eqref{eq:polarization} must have consistent phases before averaging over $\bm{k}_{\alpha}^{\perp}$. To ensure that no random branch jumps occur during the propagation, we match the string phases at times $t$ and $t + \Delta t$ at each time step. \\

%%%%%%%%%%%%%%%%%%%%%%%%
\section{Results and discussion}\label{section:results}
%%%%%%%%%%%%%%%%%%%%%%%%

\subsection{Ground state and linear response}

\begin{figure}
\center{\includegraphics[width=0.8\linewidth]{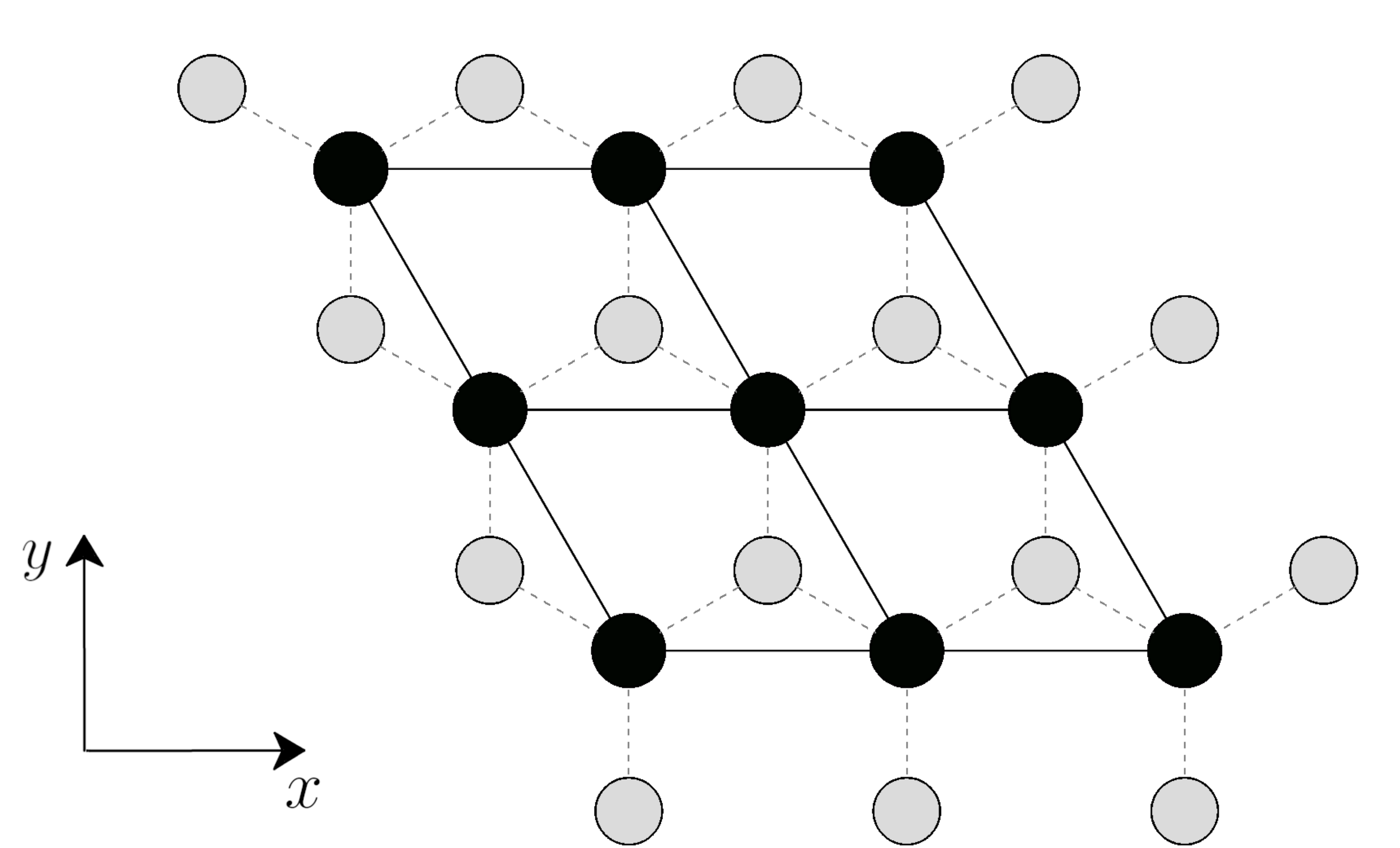}}
\caption{Scheme of the considered orientation of the honeycomb h-BN structure in real space. Unit cells are shown with solid black lines. Black circles represent B atoms, and gray circles represent N atoms.}
\label{fig:structure}
\end{figure}

  For both h-BN and GeS crystals, we start with the calculation of the ground KS state with the local DFT exchange-correlation functional $V^{\rm xc}[\rho]$ within the Perdew-Burke-Ernzerhof generalized gradient approximation revised for solids (GGA-PBEsol)~\cite{perdew_prl_2008}. We then obtain the macroscopic dielectric function $\varepsilon_{ij}(\omega) = \delta_{ij} + 4 \pi P_i(\omega) / E_j(\omega)$ by performing a real-time simulation. The external field $\bm{E}(t) = E(t) \bm{e}_j$ is chosen to be the Dirac delta function $E(t) = E_0 \delta(t - t_0)$, so that the pulse is flat in the frequency space, $E(\omega) = {\rm const}$. At the postprocessing stage, we include the broadening parameter $\omega_{\rm cut}$ = 0.1 eV when evaluating the Fourier transform of the time-dependent polarization, $\bm{P}(\omega) = \int e^{i \left (\omega - i \omega_{\rm cut} \right) t} \bm{P}(t) dt$, to mimic the experimental spectral broadening.

\subsubsection{h-BN monolayer}
Figure~\ref{fig:structure} shows the structure and considered orientation of the h-BN monolayer. The two-dimensional unit cell vectors are $a \cdot (1, \; 0)$ and $a \cdot (-1/2, \; \sqrt{3}/2)$, where the lattice constant and the nearest-neighbor distance between atoms of the same species is $a$~=~2.51~\r{A} ($\sim$4.74~a.u.).

\begin{figure*}
%\centering
\subfloat[\label{fig:bdos_hbn}]{\includegraphics[width=0.45\textwidth]{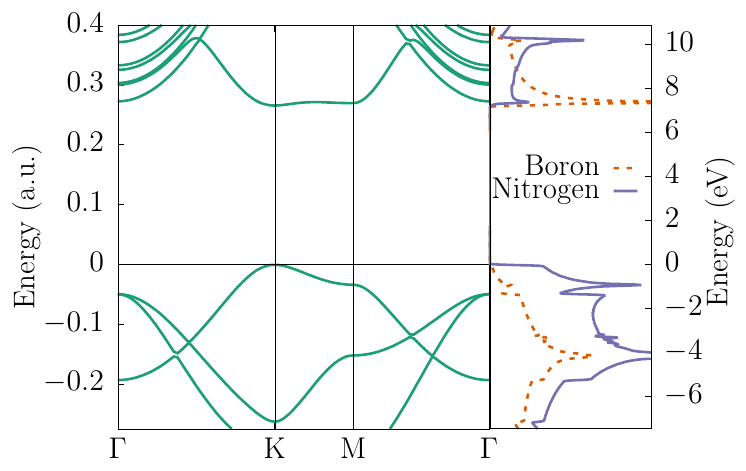}}
\quad
\subfloat[\label{fig:linear_hbn}]{\includegraphics[width=0.45\textwidth]{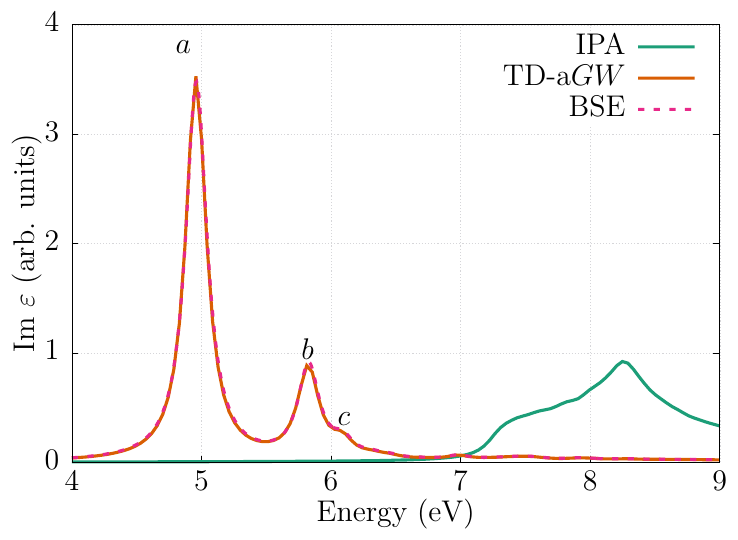}}
\caption{(a) Band structure and partial density of states of the h-BN monolayer in the ground state obtained within the DFT framework. The conduction band is shifted 2.62 eV upwards via the scissor operator~\eqref{eq:scissor} to mimic the $GW$ quasiparticle gap. (b) Imaginary part of the dielectric function Im $\varepsilon$ of an h-BN monolayer calculated with time-domain response to the delta-shaped pulse. The IPA curve was obtained within the independent particle approximation (Eq.~\eqref{eq:hamIPA}), while the TD-a$GW$ curve is obtained with the full Hamiltonian~\eqref{eq:hamFull}. The dashed BSE curve is obtained with the BSE. Lorentzian broadening of 0.1~eV is included in the postprocessing-stage Fourier transform.}
\end{figure*}

 We choose a two-dimensional $\bm{k}$ grid of 60 $\times$ 60 points equally distributed over the Brillouin zone. To mimic the direct $GW$ quasiparticle gap of 7.25 eV~\cite{paleari_2dmats_2018} located at the K point, we apply a quasiparticle scissor correction as defined in Eq.~\eqref{eq:scissor} of $\Delta E^{\rm QP} = $ 2.62 eV.  Figure~\ref{fig:bdos_hbn} shows the band structure and partial density of states (PDOS) obtained with the ground state Hamiltonian~\eqref{eq:gs_ham}. Near the gap, valence states are essentially concentrated on N sites, which agrees with the literature~\cite{galvani_prb_2016, sangalli_prb_2023}.  \\

\begin{comment}

\begin{figure}[t]
\center{\includegraphics[width=0.95\linewidth]{bands_with_dos_hbn.pdf}}
\caption{Band structure and partial density of states of the h-BN monolayer in the ground state obtained within the DFT framework. The conduction band is shifted 2.62 eV upwards via the scissor operator~\eqref{eq:scissor} to mimic the $GW$ quasiparticle gap.}
\label{fig:bdos_hbn}
\end{figure}
\end{comment}
 Only the lowest CB states and the highest-lying VB states are expected to be involved in the dynamics with the laser field parameters we are using. Therefore, we set the number of states in the CB $n_{\rm empty}$ = 2 and consider two highest-lying VB states for each $\bm{k}$ point, when performing the time propagation. In the actual calculations, only the two highest-lying states are evolved in time ($n_{\rm active} = $ 2), while the other states remain fixed at their initial values. \\ 

 We present the simulation results for $\Im\varepsilon(\omega) $ in Fig.~\ref{fig:linear_hbn}, where $\varepsilon_{xx}(\omega) = \varepsilon_{yy}(\omega) \equiv \varepsilon(\omega)$ due to the symmetry of the system. Three excitonic peaks (marked by $a$, $b$ and $c$) are clearly visible within the gap defined by the scissor-shifted IPA-level response. Based on the symmetries of the corresponding excitonic wavefunctions obtained with the solutions of the BSE, these peaks can be attributed to the $1s$-, $2p$- and $2s$-like states, respectively~\cite{galvani_prb_2016, paleari_2dmats_2018, zhang_prl_2022, malakhov_commphys_2024}. We note that while the shape of the peaks is fully converged, a finer $\bm{k}$-grid is required to determine peak positions with more than about 0.3~eV accuracy achieved here. However, these positions can be adjusted using the scissor operator. The point symmetry group of h-BN is $D_{3h}$ with six irreducible representations: $A_1'$, $A_2'$, $E'$, $A_1''$, $A_2''$ and $E''$. Since the in-plane electric dipole transition operator belongs to $E'$, only $E'$ states are visible in the linear response~\cite{dresselhaus_group_2007, attaccalite_prb_2018}, and two additional excitonic $2p$-like states belonging to $A_2'$ and $A_1'$ are dark~\cite{galvani_prb_2016}.

\begin{comment}
\begin{figure}[h]
\center{\includegraphics[width=0.95\linewidth]{linear_hbn.pdf}}
\caption{Imaginary part of the dielectric function Im $\varepsilon$ of an h-BN monolayer calculated with time-domain response to the delta-shaped pulse. (IPA) curve was obtained within the independent particle approximation (Eq.~\eqref{eq:hamIPA}), while the dashed curve (TD-a$GW$) is obtained with the full Hamiltonian~\eqref{eq:hamFull}. (BSE) curve is obtained with the BSE. Lorentzian broadening of 0.1~eV is included in the postprocessing-stage Fourier transform.}
\label{fig:linear_hbn}
\end{figure}
\end{comment}

We also compare the linear response dielectric function with the one obtained from the direct solution of the BSE using the \texttt{exciting} code. As expected~\cite{attaccalite_prb_2011}, the BSE curve is almost indistinguishable from the TD-a$GW$ one.

\begin{figure}[b]
\center{\includegraphics[width=0.6\linewidth]{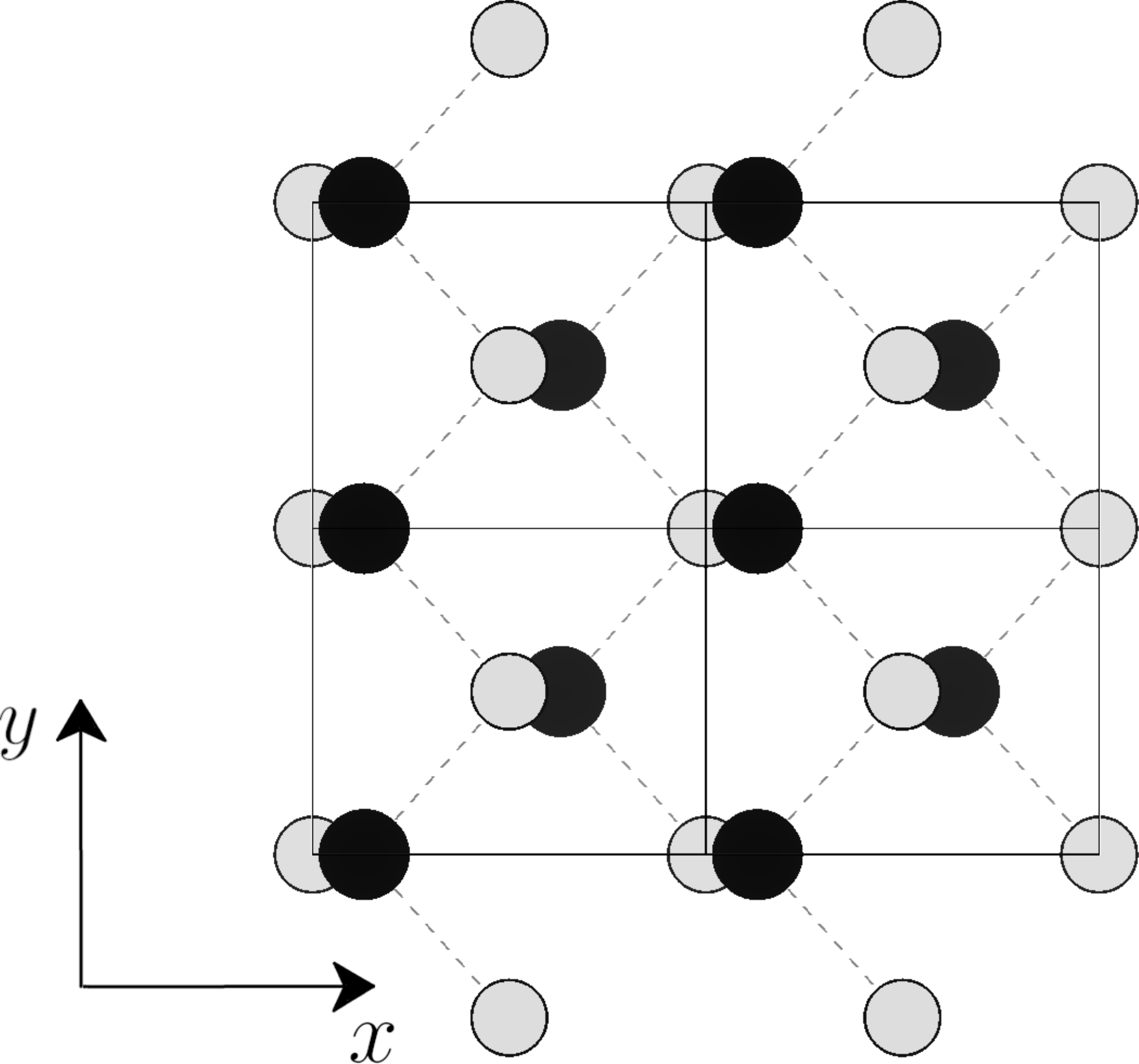}}
\caption{Scheme of the considered orientation of the GeS monolayer in real space. Unit cells are shown with solid black lines. Black circles represent Ge atoms, and gray circles represent S atoms. Direction $x$ is the ``armchair'' direction, and $y$ is the ``zigzag'' one.}
\label{fig:structure_ges}
\end{figure}

\subsubsection{GeS monolayer}

\begin{figure*}
%\centering
\subfloat[\label{fig:bdos_ges}]{\includegraphics[width=0.45\textwidth]{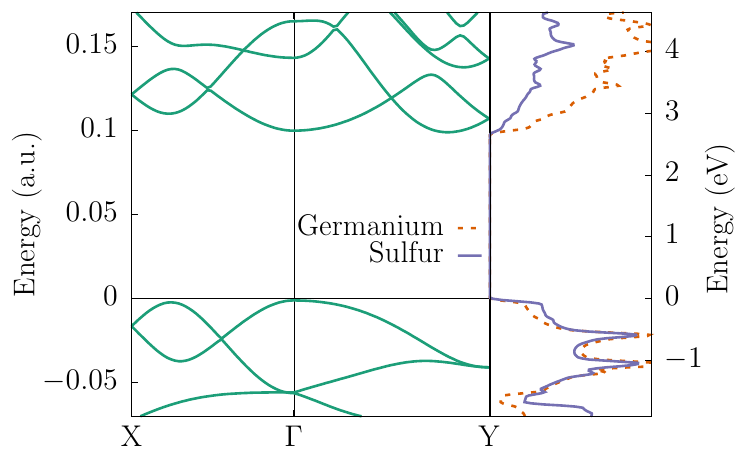}}
\quad
\subfloat[\label{fig:linear_ges}]{\includegraphics[width=0.45\textwidth]{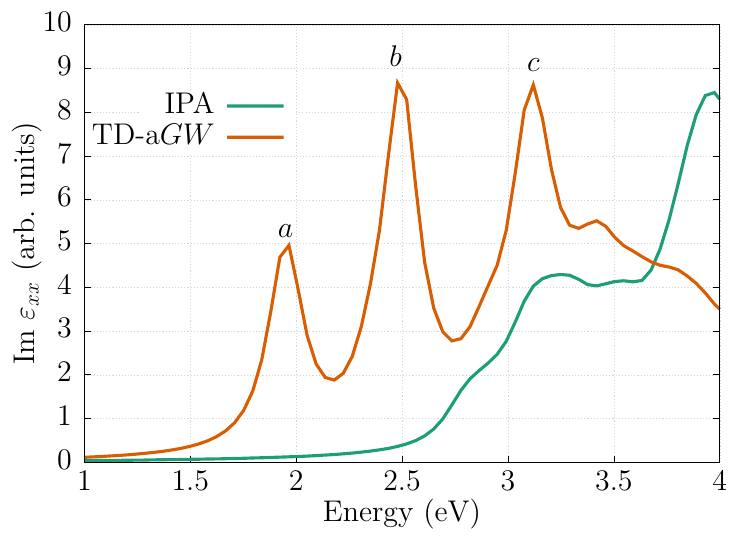}}
\caption{(a) Band structure and partial density of states of the GeS in the ground state obtained within the DFT framework. The conduction band is shifted by 1.07 eV upwards via the scissor operator~\eqref{eq:scissor} to mimic the $GW$ quasiparticle band gap. (b) Imaginary part of the dielectric function Im $\varepsilon_{xx}$ of a GeS monolayer calculated with time-domain response to the delta-shaped pulse. The IPA curve was obtained within the independent particle approximation (Eq.~\eqref{eq:hamIPA}), while the TD-a$GW$ curve is obtained with the full Hamiltonian~\eqref{eq:hamFull}. Lorentzian broadening of 0.1~eV is included in the postprocessing-stage Fourier transform. Direction $x$ is the ``armchair'' direction of a GeS monolayer.}
\end{figure*}

Figure~\ref{fig:structure_ges} shows the structure and the considered orientation of the GeS monolayer. The unit cell is formed by the orthogonal vectors $a_x \cdot (1, \; 0)$ and $a_y \cdot (0, \; 1)$ with the lattice constants $a_x$~=~4.43~\r{A} ($\sim$8.37~a.u.) and $a_y$~=~3.67~\r{A} ($\sim$6.94~a.u.). Atom positions in the lattice coordinates are set to (0.631, 0.5, 0.986) and (0.131, 0, 0.120) for Ge; and (0, 0, 0) and (0.5, 0.5, 0.106) for S. The geometry parameters are taken from the Ref.~\cite{esteve_npj_2025}. For the GeS monolayer, we use 40$\times$40 $\bm{k}$ grid. A scissor shift $\Delta E^{\rm QP}$ is set to 1.07 eV to mimic the quasiparticle gap of 2.72~eV~\cite{esteve_npj_2025}. The calculated band structure and the corresponding PDOS are shown in Fig.~\ref{fig:bdos_ges}. Since the PDOS in the considered energy range is nearly identical for atoms of the same species, the displayed PDOS is summed over all atoms of each species. \\

\begin{comment}
\begin{figure}[b]
\center{\includegraphics[width=0.95\linewidth]{bands_with_dos_ges.pdf}}
\caption{Band structure and partial density of states of the GeS in the ground state obtained within the DFT framework. The conduction band is shifted by 1.07 eV upwards via the scissor operator~\eqref{eq:scissor} to mimic the $GW$ quasiparticle band gap.}
\label{fig:bdos_ges}
\end{figure}
\end{comment}

In the real-time simulations, we only evolve two highest-lying VB states, and choose $n_{\rm empty} = 2$ in the CB. Figure~\ref{fig:linear_ges} we shows the linear response spectrum with three clearly visible excitonic peaks.

\begin{comment}
\begin{figure}[h]
\center{\includegraphics[width=0.95\linewidth]{linear_ges.pdf}}
\caption{Imaginary part of the dielectric function Im $\varepsilon_{xx}$ of a GeS monolayer calculated with time-domain response to the delta-shaped pulse. (IPA) curve was obtained within the independent particle approximation (Eq.~\eqref{eq:hamIPA}), while the dashed curve (TD-a$GW$) is obtained with the full Hamiltonian~\eqref{eq:hamFull}. Lorentzian broadening of 0.1~eV is included in the postprocessing-stage Fourier transform. Direction $x$ is the ``armchair'' direction of a GeS monolayer.}
\label{fig:linear_ges}
\end{figure}
\end{comment}

\subsection{Real-time charge dynamics}
For the real-time calculations, we assume a laser pulse  with a $\sin^2$-shaped time-dependent vector potential and different other parameters which are specified below. \update{All the quantities are plotted in arbitrary units to emphasize qualitative trends rather than quantitative behavior.} \\

\subsubsection{h-BN monolayer}

The authors of Ref.~\cite{malakhov_commphys_2024} studied a light-induced charge migration in a scenario where both $2s$ and $2p$ excitons are excited by a single circularly-polarized pulse, using the tight-binding (TB) model. They showed that such excitation results in quantum beats in the electric current, corresponding to the charge migration between $2s$ and $2p$ excitonic states in both momentum and coordinate spaces. To see our implementation in action, and to verify whether this behavior persists beyond the TB simulations, and under less symmetrical (i.e. linearly-polarized) perturbation, we set the laser pulse intensity to approximately 10$^{10}$~W/cm$^2$, pulse duration to approximately 20 fs, and the central photon energy to $\approx$ 6.12 eV to excite both $b$ and $c$ excitons shown in Fig.~\ref{fig:bdos_hbn}. \\

\begin{figure*}
\centering
\subfloat[\label{fig:beats}]{\parbox[t][6cm][b]{.45\textwidth}{\includegraphics[width=0.45\textwidth]{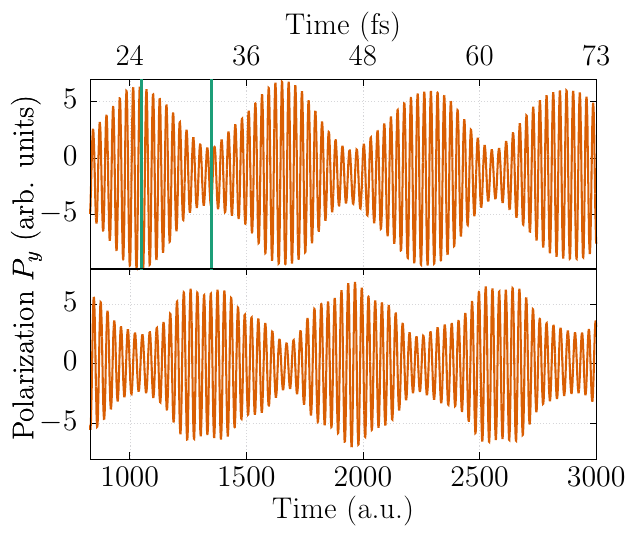}}}
\quad
\subfloat[\label{fig:beats_k}]{\parbox[t][5.5cm][b]{.45\textwidth}{\includegraphics[width=0.45\textwidth]{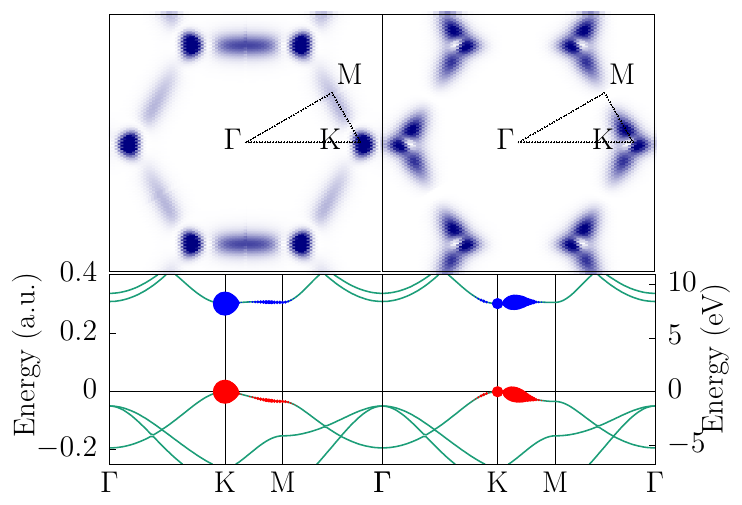}}}
\caption{(a) Time-dependent macroscopic residual polarization $P_y$ in an h-BN monolayer after excitation by a linearly-polarized, $\sin^2$-shaped pulse with a central photon energy of 6.12 eV (top panel), 3.06 eV (bottom panel), and a duration of 830 a.u. ($\approx$~21 fs). On the top panel, vertical dashed lines mark the moments in time, when the time-dependent excitonic state is almost $2s$-like (left line), and $2p$-like (right line). (b) Top row: conduction band occupations in an h-BN monolayer after excitation by the short linearly-polarized pulse with a central photon energy 6.12 eV at $t \approx$ 25~fs (left) and $t \approx$~32.5~fs (right) in reciprocal space.  The selected time points correspond to the vertical lines in Fig.~\ref{fig:beats}. Bottom row: change in the occupations of unperturbed states along the standard $\bm{k}$-path at the same times. Red and blue colors indicate the positive and negative charge differences, respectively, while point size represents the magnitude of the change.}
\end{figure*}

In Fig.~\ref{fig:beats} (top panel), we show our results for the time-dependent residual  macroscopic polarization obtained  using   Eq.~\eqref{eq:polarization}. One can clearly see the quantum beats pattern with a period of approximately 16 fs, related to the relative positions of the excitonic peaks $b$ and $c$ as $2 \pi / 16$~fs $\approx 0.25$~eV. \\

In the top row of Fig.~\ref{fig:beats_k}, we present the $\bm{k}$-space CB occupations at the time moments indicated by the vertical lines in Fig.~\ref{fig:beats}, calculated as
\begin{equation}
    \rho_{\bm{k}}(t) = \sum_{m = 1}^{n_{\rm occupied}} \sum_{n \in {\rm CB}} \left| \overlap{\phi_{\bm{k}m} (\bm{r}, t)}{\phi_{\bm{k}n} (\bm{r}, t = 0)} \right |^2.
\end{equation}
\update{We note that the times can be chosen differently as long as they correspond to a peak-node pair in the beat pattern.} Based on the analysis of the BSE solutions obtained in Ref.~\cite{galvani_prb_2016}, the left and the right distributions can be attributed to the $2s$ and $2p$-$E'$ excitonic states, respectively, with the symmetry being partially broken by the $y$-polarized laser pulse. For completeness, the bottom row of Fig.~\ref{fig:beats_k} shows the change in the state occupation relative to the ground state along the standard $\bm{k}$-path. We also observe charge migration in and out of the K point reported in Ref.~\cite{malakhov_commphys_2024}. Since the change in the charge density in real space involves integral contributions from all unit cells, the resulting distributions for both excitonic states look qualitatively the same: charge density oscillates between the B and N sites. Therefore, the real-space charge density plots are not shown here.

\begin{comment}
\begin{figure}[t]
\center{\includegraphics[width=0.95\linewidth]{beats_hbn.pdf}}
\caption{Time-dependent macroscopic residual polarization $P_y$ in an h-BN monolayer after excitation by a linearly-polarized, $\sin^2$-shaped pulse with a central photon energy of 6.12 eV (top panel), 3.06 eV (bottom panel), and a duration of 830 a.u. ($\approx$~21 fs). On the top panel, vertical dashed lines mark the moments in time, when the time-dependent excitonic state is almost $2s$-like (left line), and $2p$-like (right line).}
\label{fig:beats}
\end{figure}

\begin{figure}[h]
\center{\includegraphics[width=0.95\linewidth]{kmap_hbn.pdf}}
\caption{Top row: conduction band occupations in an h-BN monolayer after excitation by the short linearly-polarized pulse with a central photon energy 6.12 eV at $t \approx$ 25~fs (left) and $t \approx$~32.5~fs (right) in reciprocal space. Bottom row: change in the occupations of unperturbed states along the standard $\bm{k}$-path at the same times. Red and blue colors indicate the positive and negative charge differences, respectively, while point size represents the magnitude of the change. The selected time points correspond to the vertical lines in Fig.~\ref{fig:beats}.}
\label{fig:beats_k}
\end{figure}
\end{comment}

\begin{figure}
\center{\includegraphics[width=0.95\linewidth]{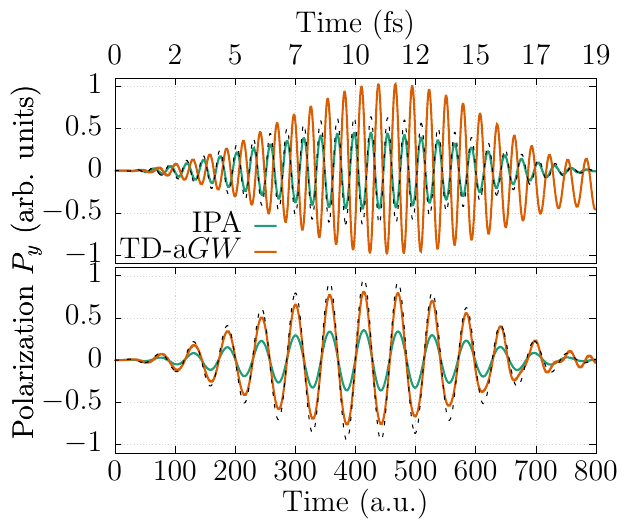}}
\caption{Time-dependent macroscopic polarization $P_y$ in an h-BN monolayer during excitation by a linearly-polarized, $\sin^2$-shaped pulse with a central photon energy of 6.12 eV (top panel), 3.06 eV (bottom panel), and a duration of 830 a.u. ($\approx$~21 fs). (IPA) curve was obtained within the independent particle approximation (Eq.~\eqref{eq:hamIPA}), (TD-a$GW$) curve corresponds to the calculations with dynamical electron-hole interactions (Eq.~\eqref{eq:hamFull}). Dashed line depicts the electric field strength. }
\label{fig:polarization_field}
\end{figure}

To include additional excitonic states in  charge migration dynamics, we change the central photon energy of the laser field from 6.12 eV to 3.06 eV. Two photons are now necessary to excite the system resulting in a change in the selection rules. The two-photon process can also trigger the formation of the $A_1'$ exciton~\cite{dresselhaus_group_2007}. Since the two-photon process is much less probable than the one-photon one, we adjust the field peak amplitude to 10$^{11}$~W/cm$^2$ to achieve the CB population comparable to the previous case of around 10$^{-3}$ excitations per unit cell. \\

The bottom panel of Fig.~\ref{fig:beats}  shows the macroscopic polarization after a two-photon excitation. Although the transition energy is the same, the charge dynamics now involve more than two excitonic states, which distorts the beats pattern. We also observe that the phase of the polarization beating is shifted by $\pi/2$ relative to the linear-response one. Since the $2p$-$A_1'$ exciton appears nearly identical to the $2p$-$E'$ exciton in reciprocal space,~\cite{galvani_prb_2016} the occupations presented in Fig.~\ref{fig:beats_k} stay qualitatively the same in the two-photon scenario. \\

\begin{figure*}[]
\subfloat[\label{fig:pol_ges}]{\parbox[t][6cm][b]{.45\textwidth}{\includegraphics[width=0.45\textwidth]{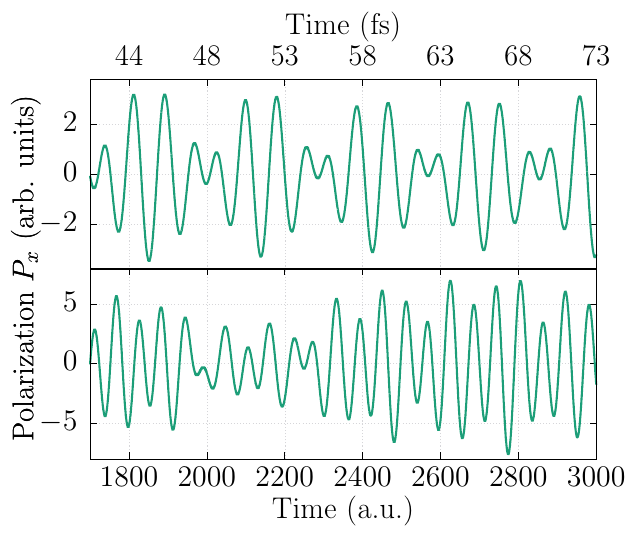}}}
\quad
\subfloat[\label{fig:ges_k}]{\parbox[t][5.5cm][b]{.45\textwidth}{\includegraphics[width=0.45\textwidth]{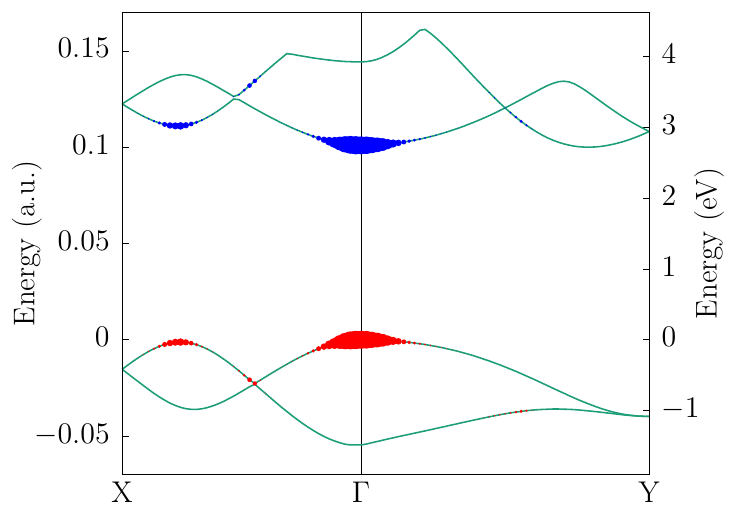}}}
\caption{(a) Time-dependent \update{residual} macroscopic polarization $P_x$ in a GeS monolayer after excitation by a linearly-polarized, $\sin^2$-shaped pulse with a central photon energy of 2.01 eV (top panel), 1 eV (bottom panel), and a duration of 1700 a.u. ($\approx$~41 fs). Direction $x$ is the ``armchair'' direction of a GeS monolayer. (b) Change in the occupations of unperturbed states along the standard $\bm{k}$-path in an GeS monolayer after excitation by the short linearly-polarized pulse with a central photon energy 2.01 eV at $t \approx$ 48~fs. Red and blue colors indicate the positive and negative occupation differences, respectively, while point size represents the magnitude of the change.}
\end{figure*}

Finally, we show the ``in-field'' macroscopic polarization response in Fig.~\ref{fig:polarization_field}. The top panel corresponds to the one-photon case, while the bottom panel shows the two-photon response. We obtain the polarization both with (TD-a$GW$) and without (IPA) the self energy term, i.e. including and excluding electron-hole interactions. Since the central photon energy of the field is set below the IPA energy gap, transition probability is negligible within the independent particle picture. Thus,  polarization follows the external field in both the linear and nonlinear regimes within this approximation. Due to excitonic effects, the transition probability is significant at a photon energy of 6.12 eV as can be seen in Fig.~\ref{fig:linear_hbn}. As a result, excitonic effects significantly impact the system's response in the linear regime. Even when the field is present, the time-dependent polarization is dominated by the contribution due to electronic excitations. The non-zero imaginary part of the dielectric function naturally manifests as a shift in oscillation phase. In the two-photon case, excitonic effects are naturally suppressed due to the dominance of the linear response.

\subsubsection{GeS monolayer}

In the case of a GeS monolayer, we set the laser pulse duration to approximately 41 fs. The central photon energy is set to 2.01 eV in the linear regime and to 1 eV in the nonlinear regime so thatthe two excitonic states marked by $a$ and $b$ in the Fig.~\ref{fig:linear_ges} are excited. We choose the longer pulse to narrow the electric field's Fourier transform in the frequency space, which better isolates the desired excitation. \\

\update{
The residual macroscopic polarization is shown in the Fig.~\ref{fig:pol_ges}. As in the previous case, the clear periodic pattern produced by the single-photon excitation is destroyed due to the excitation of additional states in the two-photon process. To understand this behavior, we refer to the exciton structure of GeS presented in Fig.~3 of Ref.~\cite{dien_prb_2023} that was calculated with the BSE. In Ref.~\cite{dien_prb_2023}, the exciton $a$ is denoted as I and the exciton $b$ is denoted as III. It follows from these results that while only excitons $a$ and $b$ (see Fig.~\ref{fig:linear_ges}) are active in the linear response regime for the photon energy we consider, there is a number of dark excitonic states in the vicinity of $b$. Transitions to these dark states can be triggered in the two-photon process. \\

}

\begin{comment}
\begin{figure}[!tbh]
\center{\includegraphics[width=0.95\linewidth]{beats_ges.pdf}}
\caption{Time-dependent macroscopic polarization $P_x$ in a GeS monolayer after excitation by a linearly-polarized, $\sin^2$-shaped pulse with a central photon energy of 2.01 eV (top panel), 1 eV (bottom panel), and a duration of 1700 a.u. ($\approx$~41 fs). Direction $x$ is the ``armchair'' direction of a GeS monolayer.}
\label{fig:pol_ges}
\end{figure}
\end{comment}

Occupations along the standard $\bm{k}$ path at \update{$t \approx$ 48 fs} are shown in Fig.~\ref{fig:ges_k}. Since the energy difference between the excitonic states is comparable to their energy difference from the ground state, the time evolution does not exhibit beatings. Therefore, the moments, when only one of the $a$ or $b$ excitonic states contributes are not distinguishable. Instead, we observe the ``trembling'' distribution, in which both states are always present.

\begin{comment}
\begin{figure}[!tbh]
\center{\includegraphics[width=0.95\linewidth]{occs_ges.pdf}}
\caption{Change in the occupations of unperturbed states along the standard $\bm{k}$-path in an GeS monolayer after excitation by the short linearly-polarized pulse with a central photon energy 2.01 eV at $t \approx$ 48~fs. Red and blue colors indicate the positive and negative occupation differences, respectively, while point size represents the magnitude of the change.}
\label{fig:ges_k}
\end{figure}
\end{comment}

%%%%%%%%%%%%%%%%%%%%%%%%
\vspace{1em}
\section{Summary}\label{section:summary}
%%%%%%%%%%%%%%%%%%%%%%%%
We investigated exciton dynamics in two-dimensional h-BN and GeS monolayers exposed to an ultrashort laser pulse, using the RT-a$GW$ approximation, capturing electron–hole correlations beyond the independent-particle picture. We implemented a highly accurate real-time propagation scheme based on the dynamical Berry phase approach to treat the coupling to time-dependent electric fields in the all-electron \texttt{exciting} package, employing the LAPW+lo basis set. Electron-hole interactions are incorporated through a non-local self-energy operator derived from the MBPT, which is equivalent to the BSE in the weak-field regime. Using our implementation, we studied response of the two-dimensional materials to an external field in linear and nonlinear regimes. We analyzed the impact of electron-hole interactions on electron dynamics during the action of a pump pulse. We demonstrated that they can result in different amplitudes and shifted phases of polarization compared to polarization calculated within the IPA. We also demonstrated the emergence of a quantum-beat scenario, which can be suppressed by breaking the symmetry of the driving field.
%%%%%%%%%%%%%%%%%%%%%%%%
\vspace{1em}
\section*{Acknowledgments}
This work has been supported by the Volkswagen Foundation grant number 96237.
%%%%%%%%%%%%%%%%%%%%%%%%

%%%%%%%%%%%%%%%%%%%%%%%%

\bibliography{refs.bib}

\end{document}